\documentclass[prd,aps,twocolumn,showpacs]{revtex4}

\def\P{{\cal P}}

\def\be{\begin{equation}}
\def\ee{\end{equation}}
\def\bea{\begin{eqnarray}}
\def\eea{\end{eqnarray}}

\newcommand\lsim{\mathrel{\rlap{\lower4pt\hbox{\hskip1pt$\sim$}}
    \raise1pt\hbox{$<$}}}
\newcommand\gsim{\mathrel{\rlap{\lower4pt\hbox{\hskip1pt$\sim$}}
    \raise1pt\hbox{$>$}}}

\def\calH{{\cal H}}

\newcommand\bfk{{\bf k}}

\newcommand\bfp{{\bf p}}
\newcommand\bfq{{\bf q}}

\newcommand\bfx{{\bf x}}

\newcommand\os{\overline{\sigma}}

\begin{document}
\draft

%
%
\input epsf
\renewcommand{\topfraction}{0.99}
\renewcommand{\bottomfraction}{0.99}

\title{The Maximal Amount of Gravitational Waves in the Curvaton Scenario}
\author{N. Bartolo$^1$, S. Matarrese$^1$, A. Riotto$^{1,2}$ and 
A. V\"aihk\"onen$^{2}$}
\address{(1) Department of Physics and INFN
Sezione di Padova, via Marzolo 8, I-35131 Padova, Italy}
\address{(2) D\'epartement de Physique Th\'eorique, Universit\'e de
Gen\`eve 4, 24 Quai Ansermet, Gen\`eve 1211, Switzerland}

\date{\today} 
\pacs{98.80.cq \hfill DFPD 07/A/08}

\begin{abstract}
\noindent
The curvaton scenario for the generation of the cosmological curvature 
perturbation on large scales represents an alternative to the standard 
slow-roll scenario of inflation in which the observed density perturbations 
are due to fluctuations of the inflaton field itself. 
Its basic assumption is that the initial curvature perturbation 
due to the inflaton field is negligible. This is  attained by 
lowering the energy scale of inflation, thereby highly suppressing 
the amount of gravitational waves produced during inflation. 
We compute the power-spectrum of the gravitational waves generated 
at second order in perturbation theory by the curvaton (isocurvature) 
perturbations between the end of inflation and the curvaton decay. 
An interesting property of this contribution to the tensor 
perturbations is that it is directly proportional to the amount 
of non-Gaussianity predicted within the curvaton scenario. 
We show that the  spectrum of gravitational waves may be in the range of 
future gravitational wave detectors.

\end{abstract}


\maketitle
\noindent
\noindent
Inflation \cite{guth81,lrreview} has  become the dominant 
paradigm for understanding the 
initial conditions for structure formation and for Cosmic
Microwave Background (CMB) anisotropy. In the
inflationary picture, primordial density and gravity-wave fluctuations are
created from quantum fluctuations ``redshifted'' out of the horizon during an
early period of superluminal expansion of the universe, where they
are ``frozen'' \cite{muk81,bardeen83}. 
These 
perturbations at the surface of last scattering are observable as temperature 
anisotropy in the CMB. 
The last and most impressive confirmation of the inflationary paradigm has 
been recently provided by the data 
of the Wilkinson Microwave Anisotropy Probe (WMAP) mission which has 
marked the beginning of the precision era of CMB measurements in space
\cite{wmap3}.

Despite the simplicity of the inflationary paradigm, the mechanism
by which  cosmological adiabatic perturbations are generated  is not
yet established. In the standard slow-roll scenario associated
to single-field models of inflation, the observed density 
perturbations are due to fluctuations of the inflaton field itself when it
slowly rolls down along its potential. 
When inflation ends, the inflaton $\phi$ oscillates about the minimum of its
potential $V(\phi)$  and decays, thereby reheating the universe. 
As a result of the 
fluctuations
each region of the universe goes through the same history but at slightly
different times. The 
final temperature anisotropies are caused by the fact that
inflation lasts different amounts of time in different regions of the universe
leading to adiabatic perturbations. 

An alternative to the standard scenario is represented by the curvaton 
mechanism
\cite{curvaton,LUW} where the final curvature perturbations
are produced from an initial isocurvature perturbation associated to the
quantum fluctuations of a light scalar field (other than the inflaton), 
the curvaton, whose energy density is negligible during inflation. The 
curvaton isocurvature perturbations are transformed into adiabatic
ones when the curvaton decays into radiation much after the end 
of inflation. 

Contrary to the standard picture, the curvaton  
mechanism exploits the fact that 
the total curvature perturbation (on uniform density hypersurfaces)
$\zeta$ can change on arbitrarily large scales due to a non-adiabatic
pressure perturbation    which may be 
present  in a multi-fluid system.
While the entropy
perturbations evolve independently of the curvature perturbation on
large scales,  the evolution of the large-scale curvature is
sourced by entropy perturbations. 

During inflation, the curvaton energy density is 
negligible and isocurvature perturbations with
a flat spectrum are produced in the curvaton field $\sigma$,

\begin{equation}
\delta\sigma_\bfk = \left(\frac{H_*}{2\pi}\right),
\end{equation}
where $H_*$ is the value of the Hubble rate during inflation. 
After the end
of inflation, 
the curvaton field oscillates during some radiation-dominated era,
causing the ratio  between its energy density and the radiation
energy density to grow, 
thereby converting the initial isocurvature into curvature 
perturbation. The  energy density $\rho_\sigma$
will then be proportional to the square of the oscillation amplitude,
and will scale like the inverse of the locally-defined comoving volume
corresponding to matter domination. On the spatially flat slicing, 
corresponding to uniform local expansion, its perturbation has a constant
value $
\delta\rho_\sigma/\rho_\sigma \simeq \left(\delta\sigma_\bfk/\os_* 
\right)$, where $\os_*$ is the value of the classical curvaton field
during inflation.

The curvature perturbation $\zeta$ is supposed to be negligible when the
curvaton starts to oscillate, growing during some radiation-dominated
era when $\rho_\sigma/\rho\propto a$, where $a$ is the scale factor.
After the curvaton decays $\zeta$  becomes constant. In 
the approximation that the curvaton decays instantly
it is then given by 
\begin{equation} \label{eq:curvpert}
\zeta_\bfk \simeq r \left(\frac{\delta \sigma_\bfk}{\os_*} \right),
\end{equation} 
where $r\equiv (\rho_\sigma/\rho)_{D}$ 
and the subscript $D$ denotes the epoch of decay. The corresponding spectrum
is \cite{curvaton}
\begin{equation}
\label{spectrum}
{\cal P}_\zeta^{\frac{1}{2}}\simeq
r  \left(\frac{H_*}{2\pi \os_*}\right).
\end{equation}
It is nearly scale-invariant under the approximation
that the curvaton field is effectively massless during inflation. 

The generation of gravity-wave fluctuations
is another
 generic prediction of an accelerated  de Sitter expansion of the universe.
Gravitational waves, whose possible observation might come from the  
detection of  the $B$-mode of polarization in the
CMB anisotropy \cite{polreview},   
may be viewed as  ripples of spacetime around the  background metric

\begin{equation}
g_{\mu\nu}=a^2(\tau)(d\tau^2-\left(\delta_{ij}+h_{ij}\right)
dx^i dx^j),
\end{equation} 
where $\tau$ is the conformal
time. The  
tensor $h_{ij}$ is traceless and transverse and has two degrees polarizations,
$\lambda=\pm$.
Since gravity-wave fluctuations are (nearly)
frozen on superhorizon scales,
a way of characterizing them is to compute
their spectrum on scales larger than the horizon. 
During a de-Sitter stage characterized by the Hubble rate $H_*$, the 
power-spectrum of gravity-wave modes generated during inflation  is 

\begin{equation}
{\cal P}_{h}(k)=\frac{k^3}{2\pi^2}\sum_{\lambda=\pm}\left|
h_{\bf k}
\right|^2=\frac{8}{M_p^2}\left(\frac{H_*}{2\pi}\right)^2,
\label{ww}
\end{equation}
where $M_p=(8\pi G_N)^{-1/2}\simeq 2.4\times 10^{18}$ GeV 
is the Planck scale. Detection of  the $B$-mode of polarization in the
CMB anisotropy requires $H_*\gsim 10^{12}$ GeV \cite{sel}.

What about the expected amplitude of gravity-wave fluctuations in 
the curvaton scenario? 

The curvaton  scenario 
 liberates the inflaton from the responsibility
of generating the cosmological curvature perturbation. Its
 basic assumption  is therefore that the initial
curvature perturbation due to the inflaton field is fully negligible. 
In the standard slow-roll inflationary models where the fluctuations 
of the inflaton field
$\phi$ are responsible for the curvature perturbations, the 
power-spectrum of the curvature
perturbation  is given by

\begin{equation}
{\cal P}_{\zeta}(k)=
\frac{1}{2 M_p^2\epsilon}\left(\frac{H_*}{2\pi}\right)^2
\left(\frac{k}{aH_*}\right)^{n_\zeta-1},
\label{vv}
\end{equation}
where $n_{\zeta}\simeq 1$ is the spectral index and 
$\epsilon=(\dot{\phi}^2/2M_p^2 H_*^2)$
is the standard  slow-roll parameter. 
Requiring that the contribution (\ref{vv}) is much smaller than
the value required to match the CMB anisotropy
imposes $H_*\ll 10^{-5}\,M_p$. 
This implies that the curvaton scenario
predicts
the  amplitude  of gravitational waves generated during inflation
 (\ref{ww}) far too small to be
detectable by future satellite experiments aimed at
observing the $B$-mode of the CMB polarization (see however \cite{prz}). 

This is not the full story though. 
Gravitational waves are inevitably generated at second order in 
perturbation theory by the curvature perturbations \cite{gws1,gws2,gws3}. 
This scalar-induced contribution can be computed 
directly from the observed density perturbations and 
general relativity and is, in this sense, 
independent of the cosmological model for generating the perturbations.
The generation of course takes place after the curvature perturbation
is generated. 

In the standard scenario, 
 where the curvature perturbation
is produced during inflation, the production of tensor modes
occurs after inflation 
when the curvature perturbations re-enter the horizon. 

In the curvaton scenario, the production of tensor modes through the 
curvature perturbations may occur only  after  the curvaton
decays, {\it i.e.}  after the isocurvature perturbations
get converted into curvature fluctuations. The 
energy density of gravitational waves (per logarithmic interval)
is given by

\begin{equation}
\Omega_{\rm GW}(k,\tau)=\frac{k^2}{6{\cal H}^2(\tau)}{\cal P}_{h}(k,\tau) 
\end{equation}
and the one generated by the curvature perturbations 
results to be of order  
$\Omega_{\rm GW}\simeq 10^{-20}$, for those modes that re-entered the
horizon when the universe was radiation dominated \cite{gws2,gws3}.

What we will be concerned about in this paper is the generation
of tensor modes by the curvaton perturbations between the end of 
inflation and the time of 
curvaton decay. In other words, we are interested in the tensor modes
generated at second order when the perturbations are still
of the isocurvature nature. This contribution may be larger than the one
created by the second-order curvature perturbations after the curvaton decay.
In this sense, the spectrum  of tensor modes computed in this paper
corresponds to the maximal possible amount  of gravity waves 
within the curvaton scenario. 

An interesting aspect 
is that the contribution  to the  tensor perturbations
turns out to be 
 directly proportional to the possibly large amount
of Non-Gaussianity (NG) in the CMB anisotropies 
which is predicted within the curvaton scenario \cite{ng1,komatsu}.
NG is usually 
 parametrized in terms of the  the nonlinear  parameter 
$f_{\rm NL}$ and the latter   is
predicted to be of the order of $1/r$ in the curvaton scenario;
present-day data limit $|f_{\rm NL}|$ to be smaller than about
$10^{2}$, that is $r\gsim 10^{-2}$ \cite{wmap3,leonardo}. 
Therefore, the present
observational bound on the level of NG in the CMB can already
put an upper bound on the amount of tensor modes induced by the
curvaton perturbations. This relic gravitational radiation
may be particularly relevant in view of the realization that
space-based laser interferometers, such as the Big Bang Observer (BBO) 
and the Deci-hertz Interferometer Gravitational wave Observatory
(DECIGO), 
operating in the frequency
range between $\sim$ 0.1 Hz and 1 Hz may achieve the necessary 
sensitivity \cite{gwexpt}.

The equation for the second-order gravitational waves before
curvaton decay can be written (neglecting the first-order vector 
perturbations) as
\begin{equation}
h^{\prime\prime}_{ij}+2{\cal H}h^{\prime}_{ij}-\nabla^2 h_{ij}=-4 \kappa^2
{\cal T}_{ij}^{\,\,lm} \partial_l\delta\sigma\partial_m\delta\sigma,
\label{a}
\end{equation}
where $\calH=a^\prime/a$ is the Hubble rate, the prime stands for
differentiation with respect to the conformal time
and $\kappa^2=8\pi G_N$. 
If we define the Fourier transform of the tensor perturbations as follows
\begin{equation}
h_{ij}(\bfx,\tau)=\sum_{\lambda=\pm}
\int\frac{d^3\bfk}{(2\pi)^{3/2}}e^{i\bfk\cdot\bfx}
h^{\lambda}_\bfk(\tau)e^{\lambda}_{ij}(\bfk),
\end{equation}
where the polarization tensors 

\begin{eqnarray}
e^{+}_{ij}(\bfk)&=&\frac{1}{\sqrt{2}}
(e_i(\bfk)e_j(\bfk)+ \overline{e}_i(\bfk)
\overline{e}_j(\bfk)),\nonumber\\
e^{-}_{ij}(\bfk)&=&\frac{1}{\sqrt{2}}
(e_i(\bfk)\overline{e}_j(\bfk)- \overline{e}_i(\bfk)
e_j(\bfk))
\end{eqnarray}
are expressed in terms of orthonormal basis
vectors ${\bf e}$ and  $\overline{{\bf e}}$ orthogonal to $\bfk$, the 
projector tensor in Eq. (\ref{a}) reads
\begin{equation}
{\cal T}_{ij}^{\,\,lm} =\sum_{\lambda=\pm}
\int\frac{d^3\bfk}{(2\pi)^{3/2}}e^{i\bfk\cdot\bfx}
e^{\lambda}_{ij}(\bfk)e^{\lambda\,lm}(\bfk).
\end{equation}
In Fourier space, the equation of motion for the gravitational wave amplitude
(for each polarization) then becomes
\begin{eqnarray}
&&h^{\prime\prime}_\bfk+2{\cal H}h^{\prime}_\bfk+k^2 h_\bfk =
{\cal S}(\bfk,\tau),\nonumber\\
{\cal S}(\bfk,\tau)&=& 4 \kappa^2\int\frac{d^3\bfp}{(2\pi)^{3/2}}
e^{+\,lm}(\bfk)p_l p_m\delta\sigma_\bfp(\tau)\delta\sigma_{\bfk-\bfp}(\tau).
\label{aa}
\end{eqnarray}
The solution to this equation can be easily found to be
\begin{equation}
h_\bfk(\tau)=\frac{1}{a(\tau)}\int^\tau d\tau^\prime g_\bfk(\tau^\prime,\tau)
a(\tau^\prime){\cal S}(\bfk,\tau^\prime),
\end{equation}
where $g_\bfk(\tau^\prime,\tau)$ is the appropriate Green function
either for a radiation- or a matter-dominated period.

We split the perturbations of the curvaton field into a transfer function
piece $T_\sigma(k,\tau)$ and the primordial fluctuation $\delta\sigma_\bfk$,
\begin{equation}
\delta\sigma_\bfk(\tau)=T_\sigma(k,\tau)\,\delta\sigma_\bfk
\end{equation}
with the primordial power-spectrum defined by 
\begin{equation}
\langle\delta\sigma_\bfk \delta\sigma_\bfq\rangle=
\frac{2\pi^2}{k^3}\delta(\bfk+\bfq)
\P_{\delta\sigma}(k).
\end{equation}
The power-spectrum of the second-order gravitational waves becomes
\begin{eqnarray}
&&\P_h(k,\tau) = \nonumber \\  
&&16 \kappa^4 \int_0^\infty dp\int_{-1}^{1} 
d\cos\theta\P_{\delta\sigma}(p)\P_{\delta\sigma}(|\bfk-\bfp|)
\frac{\sin^4\theta}{a^2(\tau)} \times \nonumber\\
&&\frac{k^3p^3}{|\bfk-\bfp|^3}
\left|\int^\tau d\tau^\prime a(\tau^\prime) g_\bfk(\tau^\prime,\tau)
T_\sigma(p,\tau^\prime)T_\sigma(|\bfk-\bfp|,\tau^\prime)\right|^2, \nonumber 
\\ 
&&
\label{ps}
\end{eqnarray}
where $\cos\theta=\hat{\bfk}\cdot\hat{\bfp}$.
The second-order tensor modes are generated when the various modes $k$ 
enter the 
horizon. Meanwhile,  the  production ends when the curvaton decays. 
We will assume in the following that the whole generation of tensor modes
takes place in the 
radiation-dominated epoch. This assumption is motivated by requiring that
the NG induced by the curvaton is sizeable, which requires
the curvaton energy density  not to dominate by  the time of decay. 
Since the Hubble rate is given by 
${\cal H}=1/\tau$, a given mode $k$ enters the horizon
at $\tau_k=1/k$. Indicating by $k_D$  the mode which enters
the horizon at the time of the curvaton decay, 
$k_D=a(\tau_D)\Gamma$, where $\Gamma$ is the decay rate of the
curvaton, we may write the evolution of the scale factor as

\begin{equation}
a(\tau)=\left(\frac{k_D^2}{\Gamma}\right)\,\tau.
\end{equation} 
Trading the curvaton decay rate with the
temperature at decay $T_D$, we obtain 

\begin{equation}
k_D\simeq 10^{-8}\left(\frac{T_D}{{\rm GeV}}\right)\,\, {\rm Hz}. 
\end{equation}
After the end of inflation, the zero mode $\os$ of the curvaton field
starts oscillating at $\tau_m\equiv (1/k_D)(\Gamma/m)^{1/2}$, where
$m$ is the curvaton mass. 
Let us first consider those perturbations which enter the horizon when
the zero mode $\os$ of the curvaton decay is already oscillating, that 
is $k\lsim (m/\Gamma)^{1/2}k_D$. In this range of
wavenumbers, one can show that 
the curvaton perturbations scale as the zero mode, 
$\delta\sigma_\bfk (\tau) \sim \os\sim a^{-3/2}$. This 
allows to write  

\begin{eqnarray}
\delta\sigma_\bfk(\tau)&=&
\left(\frac{\delta\sigma_\bfk}{\os_*}\right)
\left(\frac{1}{k_D\tau}\right)^{3/2}\os_D\simeq
\frac{\zeta_\bfk}{r} \left(\frac{1}{k_D\tau}\right)^{3/2}\os_D,\nonumber\\
T_\sigma(k,\tau)&=&\left(\frac{\os_D}{\os_*}\right)
\left(\frac{1}{k_D\tau}\right)^{3/2}.
\label{tranf}
\end{eqnarray}
where $\os_D$ is the value of the curvaton zero mode at the time of decay. 
If we introduce the variables $x=\left|\bfk-\bfp\right|/k$ and $y=p/k$
and uses the radiation-dominated Green function $g_\bfk(\tau^\prime,\tau)=
\sin(k(\tau-\tau^\prime))/k$, it is easy to realize that the 
main tensor mode production happens at horizon entry, that is at $\tau\simeq
\tau_k$. Therefore, the power-spectrum (\ref{ps}) computed at 
horizon entry  is 

\begin{eqnarray}
&&\P_h(k,\tau_k) \simeq 10^2 f_{\rm NL}^2\left(\frac{k}{k_D}\right)^6
\left(\frac{\Gamma}{m}\right)^4 
\int_0^\infty dy \int_{\left|1-y\right|}^{1+y}
dx \nonumber\\
&\times&\frac{y^2}{x^2}
\left(1-\frac{(1+y^2-x^2)^2}{4 y^2}\right)^2
\P_{\zeta}(kx)\P_{\zeta}(ky),\nonumber\\
&&
\label{ps1}
\end{eqnarray}
where we have made use of the relations
$r=(\kappa^2/3\Gamma^2)(m^2\os_D^2)$ and $ f_{\rm NL}\sim 1/r$. In the
curvaton scenario the resulting curvature perturbation is 
nearly scale-invariant and we can take $\P_{\zeta}\sim (5\times 10^{-5})^2$. 
The remaining integrals in (\ref{ps1}) are dominated by the momenta
for which $x\sim y\sim (m/\Gamma)^{1/2}(k_D/k)$. 
We finally obtain an energy density today
of gravitational waves given by

\begin{equation}
\Omega_{\rm GW}\simeq  10^{-15}
\left(\frac{f_{\rm NL}}{10^2}\right)^2
\left(\frac{k}{k_D}\right)^5
\left(\frac{\Gamma}{m}\right)^{7/2},
\label{result1}
\end{equation}
valid for $k_D\lsim k\lsim (m/\Gamma)^{1/2}k_D$. 

For those perturbations which enter the
horizon before the zero mode of the curvaton field
starts oscillating, that is $k\gsim (m/\Gamma)^{1/2}k_D$, the scaling is
is $\delta\sigma_\bfk\sim a^{-1}$ for $\tau_k\lsim\tau\lsim k\tau_m^2$, that 
is till $k\gsim ma$. For $ k\tau_m^2\lsim\tau\lsim \tau_D$ 
the scaling is
is $\delta\sigma_\bfk\sim a^{-3/2}$. Meanwhile, the zero mode
$\os$ remains frozen till the mass of the curvaton becomes larger than
the Hubble rate at $\tau\sim \tau_m$. Repeating the
steps leading to Eq.~(\ref{ps1}), we find at horizon entry

\begin{eqnarray}
&&\P_h(k,\tau_k) \simeq 10^2 f_{\rm NL}^2
\left(\frac{\Gamma}{m}\right)
\int_0^\infty dy
\int_{\left|1-y\right|}^{1+y}
dx\nonumber\\
&\times& \frac{1}{x^4}
\left(1-\frac{(1+y^2-x^2)^2}{4 y^2}\right)^2
\P_{\zeta}(kx)\P_{\zeta}(ky),\nonumber\\
&&
\label{ps2}
\end{eqnarray}
which corresponds to an energy density today of 
\begin{equation}
\Omega_{\rm GW}\simeq  10^{-15}
\left(\frac{f_{\rm NL}}{10^2}\right)^2
\left(\frac{\Gamma}{m}\right),
\label{result2}
\end{equation}
valid for $k\gsim  (m/\Gamma)^{1/2}k_D$. 
Finally, let us note that the redshifted
gravitational wave at later times is always larger than 
${\cal S}/k^2$. Therefore the power-spectrum of gravity waves produced by
the curvaton fluctuations is always bigger
than the one generated by the second-order  curvature
perturbations which is inevitably generated when 
the cosmological perturbations acquire their adiabatic nature.

There is  a simple physical motivation for the fact that the 
amount of gravity 
waves generated by the curvaton decay is enhanced by powers of 
$1/r$ 
with respect to the one produced by ordinary second-order curvature
perturbations. 
Indeed, being the final adiabatic perturbations generated by the curvaton 
isocurvature perturbations, the smaller the amount of curvaton energy 
density  at decay is, the larger the curvaton fluctuations have to be: 
$\delta\sigma_{\bf k}\propto \zeta_{\bf k}/r$, see Eq.\ (\ref{eq:curvpert}). 
The isocurvature perturbations giving rise to gravity waves 
are therefore parametrically larger at horizon entry than the second-order curvature perturbations.

From our findings we deduce that the amount of gravitational waves
in the perturbative regime $\Gamma\lsim m$ can be as large
as $\Omega_{\rm GW}\simeq  10^{-15}$ maximazing the observationally 
allowed NG in the CMB anisotropies. This is quite intriguing since
such a spectrum is at the range of, {\it e.g.}, BBO and DECIGO 
interferometers. To be in the right frequency range, between $10^{-1}$ and
1 Hz, one has to impose that the curvaton decays at $T_D\lsim
10^8$ GeV. Furthermore, the correlated BBO interferometer proposal claims
a sensitivity down to $\Omega_{\rm GW}\simeq  10^{-17}$. This
would require $\Gamma\gsim 10^{-2}m $ in units of 
$\left(10^2/f_{\rm NL}\right)^2$. As a final remark, we point out that
the expressions (\ref{result1}) and (\ref{result2})
are also applicable to the so-called modulated reheating scenario
in which the curvature perturbations are due
to the fluctuations of some light field parametrizing 
the  inflaton decay rate \cite{gamma}. In such a case though the NG nonlinear
parameter $f_{\rm NL}$ does not exceed  unity \cite{komatsu}. 

In summary, we have computed the 
maximal amount of tensor perturbations which may be generated within the
curvaton scenario. It is  
directly proportional to the square of the nonlinearity  parameter 
$f_{\rm NL}$ which parametrizes the NG in CMB anisotropies.
The present observational bound on the level of NG in the CMB data 
sets already an upper bound on the amount of tensor modes induced by the
curvaton perturbations. On the other hand, a possible 
future detection of a large NG signal in CMB anisotropies would suggest a 
quantity of gravity waves at range of future gravitational wave
detectors. 

AV is partially supported by the Academy of Finland. We wish to thank Ruth Durrer for useful discussions.  


\end{document}